\documentclass[letterpaper]{article} 
\usepackage{aaai2026}  
\usepackage{times}  
\usepackage{helvet}  
\usepackage{courier}  
\usepackage[hyphens]{url}  
\usepackage{graphicx} 
\usepackage{natbib}  
\usepackage{caption} 
\usepackage{booktabs}
\usepackage{enumitem}

\newcommand{\greyT}[1]{\textcolor{black!50}{#1}}
\newcommand{\pinkT}[1]{\textbf{\textcolor[RGB]{237,114,178}{#1}}}
\newcommand{\tealT}[1]{\textbf{\textcolor[RGB]{55,126,127}{#1}}}

\title{Why and How People Check Generative AI Output for Mistakes}
\author{
    Patrick Gage Kelley\textsuperscript{\rm 1},
    Derrick Feldmann\textsuperscript{\rm 2},
    Reena Jana\textsuperscript{\rm 1},
    Colleen Thompson-Kuhn\textsuperscript{\rm 2},\\
    Allison Woodruff\textsuperscript{\rm 1}
}
\affiliations{
    \textsuperscript{\rm 1}Google, United States\\
    \textsuperscript{\rm 2}Ad Council Research Institute, United States\\
    patrickgage@acm.org, dfeldmann@adcouncil.org, reenaj@google.com, ckuhn@adcouncil.org, woodruff@acm.org
}

\begin{document}

\maketitle

\begin{abstract}
Generative AI output can contain errors, such as hallucinations, non-responsive results, or otherwise inaccurate or potentially harmful content. To explore the public's emerging understanding, attitudes, and behavior regarding such mistakes, we ran an online survey in the United States with 1,503 respondents, with a representative sample of the population. We report high public awareness of generative AI mistakes. Further, many respondents report checking generative AI output, for example, by comparing results with other online resources. We conclude with guidance for explanations and in-product disclosures about generative AI mistakes.
\end{abstract}

\section{Introduction}

As the use of generative AI (GenAI) becomes increasingly widespread \cite{bick2026,chatterji2025,sajadieh2026}, its beneficial and harmful properties have been a topic of great discussion \cite{carmichael2025,kennedy2025}. In this paper, we explore one particular issue, public opinion regarding GenAI's production of errors in its output. It is well-documented that regardless of developer, geography, or technological specifics, all current GenAI models make mistakes, such as hallucinations \cite{ji2023}, non-responsive results, or otherwise inaccurate or potentially harmful content. However, it is less clear what the general public knows of this, and what if anything they do to detect or mitigate GenAI mistakes.

To explore public understanding, attitudes, and behavior regarding GenAI mistakes, we ran an online survey in the United States with 1,503 respondents, with a sample representative of the general population. We fielded the survey in May 2025; accordingly, our work documents public perception at that moment in the rapid uptake of GenAI. We report high public awareness of generative AI mistakes, with 70\% of respondents saying GenAI makes mistakes at least sometimes. Further, a majority of respondents believe GenAI at least sometimes does all of the following: gets facts wrong; makes up things that don't exist; says things that experts wouldn't agree with; makes typos and grammatical errors; and doesn't follow the instructions in the prompt. Further, respondents indicate strong interest in accuracy, with approximately 75\% of those responding to a question about checking GenAI output confirming that they check it at least some of the time, for example, by comparing GenAI results with other online resources. We also report respondent antipathy towards general warnings about GenAI mistakes, with more favorable responses to guidance to check results and provide feedback.

Our contribution is therefore a novel study of the general public's understanding, attitudes, and behavior regarding mistakes in GenAI output. In the remainder of the paper, we review relevant background, describe our methodology, present our findings, and discuss implications including guidance for disclosures about GenAI mistakes.

\section{Related Work}

\paragraph{GenAI Adoption.} As noted in Stanford's 2026 AI Index Report, ``AI adoption is spreading at historic speed'' \cite{sajadieh2026}. Bick et al. report extremely rapid uptake of GenAI tools, with nearly 40\% of the US population using GenAI as of late 2024, arguing that work adoption of GenAI has been as fast as adoption of the personal computer, and overall adoption of GenAI has been faster than the adoption of either PCs or the internet \cite{bick2026}. The Pew Research Center details ChatGPT use specifically, reporting that about one-third of US adults had ever used ChatGPT as of early 2025 \cite{sidoti2025}; Chatterji et al. report around 10\% of the world's adult population uses ChatGPT as of July 2025 \cite{chatterji2025}. More recently, Ipsos reported that the number of US respondents who say they never use AI tools fell from 26\% in November 2025 to 17\% in April 2026 \cite{ipsos2026b}.

Evidence of rapid adoption is accompanied by data about robust personal and professional use of GenAI. Chatterji et al. report strong personal and work use of GenAI globally as of July 2025, noting that writing dominates work-related ChatGPT use \cite{chatterji2025}. Earlier in GenAI adoption, the Pew Research Center reported that 63\% of US adult workers said they don't use AI much or at all in their jobs; at the same time, 40\% of workers who do use AI chatbots say they have been extremely or very helpful in increasing their speed, and 29\% say they have been extremely or very helpful in improving the quality of their work \cite{lin2025}. Despite increased public exposure to GenAI, research on the impact of GenAI errors on attitudes and behavior remains sparse \cite{mueller2025}.

\paragraph{Public Attitudes.} Public response to AI has been characterized as a mix of excitement and concern since before GenAI permeated public consciousness \cite{kelley2021}. However, nervousness about AI has increased globally since the public launch of ChatGPT \cite{carmichael2025}, and in early 2025, 34\% of respondents said they are more concerned than excited about the increased use of AI, while 42\% of respondents said they are equally concerned and excited \cite{poushter2025}. Prominent concerns include the labor market, loss of creativity/critical thinking, misinformation, and the environment \cite{carmichael2025,ipsos2026a,kennedy2025,li2026}. Negative sentiment has escalated in recent months, with ``AI backlash'' making headlines \cite{goldberg2026,mickle2026,quinnipiac2026}.

General trust in AI has long been studied, with many recent results focusing on GenAI, e.g., Tolsdorf et al. report a tendency towards binary low/high trust in GenAI \cite{tolsdorf2025}. The Pew Research Center reports that a majority of US adults have seen AI summaries in search results, but 46\% of those don't trust them \cite{eddy2025}. The hazards of over-reliance have also been extensively investigated \cite{passi2022}, emphasizing the importance of ensuring users are appropriately cognizant of potential errors.

\paragraph{AI Mistakes.} AI has long been understood to make mistakes, often resulting in consequential harm \cite{raji2022,shelby2023}. With the advent of GenAI, new types of errors have reached public consciousness, e.g., hallucinations \cite{bohannon2023}. Beyond media reports, tools such as disclosures, disclaimers, and AI incident databases promote awareness of risks, harms, and errors \cite{mcgregor2021,mitfuturetech2025}, and there is some evidence of low to moderate public awareness of visible mistakes in GenAI output. In a survey of US adults, Li et al. found that when evaluating fictionalized scenarios, respondents mentioned GenAI failures such as hallucinations (27\%) or low quality output (28\%) \cite{li2026}. Tolsdorf et al. report that while participants perceived most risks of GenAI conversational agents to be unlikely, one segment of participants was more inclined to rate unhelpful output and misinformation as likely \cite{tolsdorf2025}. The Pew Research Center reported that in an open-ended response in a survey of US adults in June 2025, 3\% of respondents said they rate the societal risks of AI as very high due to AI mistakes including hallucinations \cite{kennedy2025}, and Lean In reports that in a nationally representative US survey, 22\% of women and 17\% of men questioned whether AI is accurate \cite{leanin2026}.

\paragraph{Human-AI Verification.} Little research has been done on how people check results of GenAI in organic settings. In one rare exception, Lee et al. present a survey of knowledge workers' self-reported behaviors. In response to cognitive priming, they reported behaviors such as checking GenAI output to ensure quality, e.g., verifying information by checking referenced or independent sources, while also expressing that barriers such as a lack of time limited their ability to think critically at work \cite{lee2025}. Gu et al. report a qualitative study of data analysts, who typically used procedural and data-oriented strategies to validate output \cite{gu2024}. Perera et al. report a qualitative study of verification practices of blind users working with GenAI in spreadsheets, observing certain behaviors similar to those we report, such as cross-checking and consulting search engines, although also emphasizing that the practices are qualitatively different and more complex for blind users as compared with sighted users \cite{perera2026}. We complement these studies by investigating verification attitudes and behaviors with a general population.

A number of remediations have been proposed to improve human verification of GenAI results. AI literacy and explainability have long been recognized as key tools for public understanding \cite{long2020,kelley2023}, with Annapureddy et al. proposing competencies specific to GenAI, including the ability to evaluate GenAI output and assess whether content is accurate and relevant \cite{annapureddy2025}. Such evaluation can be cognitively demanding \cite{tankelevitch2024}, and researchers have articulated the need for tools to help humans check GenAI output \cite{gordon2023}. Several interfaces have been proposed, including leveraging warnings \cite{nahar2024}, chain-of-thought (CoT) reasoning \cite{zhou2026}, sincere apologies \cite{mahmood2022}, or a warn-verify-audit interface \cite{laban2024}, often showing improvements on measures such as error detection rates, speed, and favorability.

Overall, despite early data points, there has been limited study of public perception of mistakes in GenAI output, and even less insight regarding why and how everyday users check GenAI results.

\section{Methodology}
We deployed a survey to investigate our research questions:
\begin{itemize}[leftmargin=1.4cm]
\item[\textbf{RQ1:}]  Do people think GenAI makes mistakes?
\item[\textbf{RQ2:}] What kinds of mistakes do they think GenAI makes, and how often?
\item[\textbf{RQ3:}] Do people check GenAI results? Why and how?
\item[\textbf{RQ4:}] What explanations and in-product statements about GenAI mistakes do people find useful?
\end{itemize}

\subsection{The Questionnaire}

At the beginning of the survey, respondents were shown the following definition of GenAI:
\begin{quote}
\emph{Generative AI is a type of artificial intelligence that creates new content based on patterns it has learned from huge amounts of existing data. An example of generative AI is an advanced chatbot that answers people's questions in a conversational way. Generative AI can also produce entirely new videos or images, summarize long documents, create outlines or reports, and more.}
\end{quote}

\noindent The questionnaire had the following sections:

\begin{itemize}
\item \textbf{Demographics} -- Including age, gender, race/ethnicity, region within the US, and household income.
\item \textbf{General Attitude and Familiarity} -- We asked if respondents were familiar with GenAI; how excited, concerned, and useful it is; and how trustworthy its results are.
\item \textbf{Respondent's Use of GenAI} -- For those familiar with GenAI, we asked how regularly they use it for personal tasks and for work tasks.
\item \textbf{GenAI Mistakes} -- We asked how often GenAI makes mistakes, and which types.
\item \textbf{Explanations and In-Product Messaging} (\emph{highlighter exercise}) -- We had respondents give positive or negative feedback on statements that explain GenAI.
\end{itemize}

\noindent The full text of the questions reported in this paper is in the Appendix.\footnote{The Appendix is available in the arXiv version of this paper. Data from this survey is also used in an Ad Council Research Institute white paper~\cite{feldmann2025}.} Questions were closed-form, open-ended, and a highlighter exercise. In the latter, users were shown statements and could freely highlight any words they chose with positive or negative ``highlighters'' representing like/helpful or dislike/not helpful. The statements were hypothetical but aligned with existing approaches across the industry.

\subsection{Survey Deployment}

We surveyed 1,503 people (767 women, 730 men, 5 non-binary/gender non-conforming, and 1 Two-Spirit) in the US on their usage, attitudes, and behaviors regarding GenAI tools. The sample was representative of US Census data on age, gender, race/ethnicity, region, and household income. Full demographics are shown in the Appendix, Table 4. The online survey was fielded May 8-14, 2025 by a professional market research agency. It was administered on mobile or desktop, and it was offered in English and Spanish. Survey length was estimated at 10 minutes and took respondents a median time of 11.7 minutes. Participants were compensated at industry standard in panel currency usable for gift cards.

In response to a filtering question, 145 respondents reported they were not at all familiar with GenAI and an additional 18 reported they first heard about it during the survey. These 163 \emph{unfamiliar} respondents were not shown some questions, including those on AI use, but were asked about their broader AI attitudes and to do the highlighter exercise.

\subsection{Data Processing and Analysis}

We used an inductive approach to explore emerging themes and common patterns in the data \cite{hinkin1998}.

\paragraph{Data Cleaning and Processing.} We performed several quality checks to ensure a sound sample. We performed pre-survey quality checks, removing 338 respondents before they entered the survey. During these checks, every potential survey completion was sent through multiple automated systems, in real time, to prevent duplicates, survey-bots, click-farms, oversampled respondents, and known low-quality respondents from entering the survey. We also performed post-survey quality checks, removing 57 respondents from the final dataset for quality concerns such as speeding, straight-lining, contradictory responses, or low quality open-ended responses. Additionally, 316 respondents dropped out during the survey and were removed from the final data set; 128 of those respondents dropped out during the initial survey screens. Responses to the highlighter exercise were analyzed to identify non-overlapping phrases, and each respondent was assigned like/helpful, dislike/not helpful, or no opinion for each phrase. Verbatim responses in Spanish were translated into English for analysis, using Google Translate.

\begin{table}[t]
\centering
\small
\begin{tabular}{lr cc cc}
& & \multicolumn{4}{c}{\textit{\large Work Tasks}} \\
& & \multicolumn{2}{c}{Use} & \multicolumn{2}{c}{Non-use} \\\cmidrule{3-6}
\textit{\large Personal} & Use & \textbf{43\%} & \greyT{648} & \textbf{17\%} & \greyT{253} \\ \cmidrule{3-6}
\textit{\large Tasks} & Non-use & \textbf{3\%} & \greyT{49} & \textbf{37\%} & \greyT{553} \\
\cmidrule{3-6}
\end{tabular}
\caption{Reported use of GenAI for personal or work tasks (Q7). \emph{Use} includes those who said very often, often, sometimes, or rarely. \emph{Non-use} includes those who said not at all or don't know, plus 163 respondents unfamiliar with GenAI.}
\label{tab:use}
\end{table}

\paragraph{Analysis of Closed-Form Responses.} Quantitative analysis was performed on unweighted responses, meaning each response was treated as being of equal importance to all others. Text-based scales were treated categorically. In some cases we group categories, e.g., we group very/extremely in our odds ratios, and we group segments for GenAI use. The statistical analysis was done in Python using the pandas library, with code that was written with AI assistance and validated by one of the authors. To understand the drivers of trust in GenAI results and likelihood of mistakes we estimated multivariable logistic regression models, which we report as Adjusted Odds Ratios. We additionally calculated dominance analysis to understand the relative drivers; demographic details drove less than 4\% of the overall variability and thus are not included in the analysis in this paper.

\paragraph{Analysis of Open-Ended Responses.} Research team members reviewed verbatim responses for all three open-ended questions and discussed emergent themes \cite{beyer1997}. During this process, we saw that responses to the checking question (Q12) were focused and particularly amenable to qualitative coding. Accordingly, the last author developed a codeframe for Q12 and applied it to those responses, in consultation with the first author. As the application of the codeframe was straightforward due to brief responses and clear concepts, coding by additional reviewers was deemed unnecessary following guidance on ``ease of coding'' \cite{mcdonald2019}.

\section{Limitations}

We note several limitations of our methodology. First, it carries the standard issues attendant with survey methodology, such as the risks of poor quality translation, or respondents misunderstanding questions, satisficing \cite{holbrook2003}, or plagiarizing open-ended responses. We have worked to minimize these risks through use of open-ended questions in conjunction with closed-form questions, as well as data quality checks. Second, the survey was conducted only in the US and may not be representative of other geographies. Third, both the landscape of GenAI and public understanding of it are changing rapidly. The responses captured here reflect a moment in time, and public opinion may shift as the technology evolves.

\begin{figure*}[t]
\centering
\includegraphics[width=0.9\textwidth]{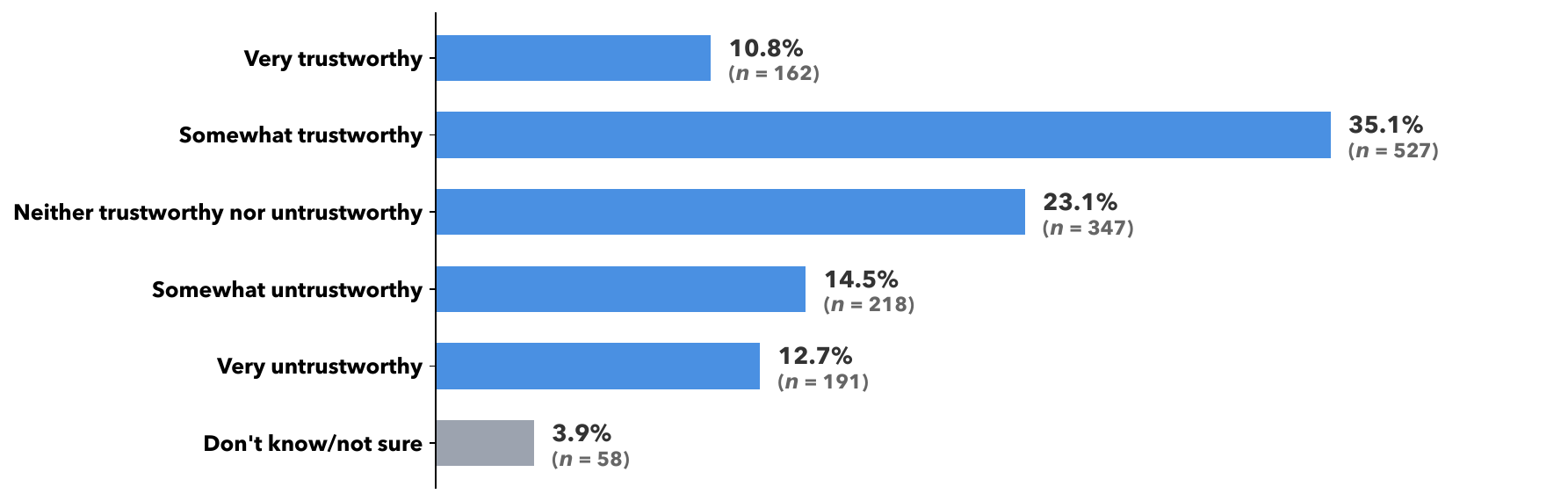}
\caption{Perceived trustworthiness of GenAI responses (Q9).}
\label{fig:trust}
\end{figure*}

\begin{figure*}[t]
\centering
\includegraphics[width=0.9\textwidth]{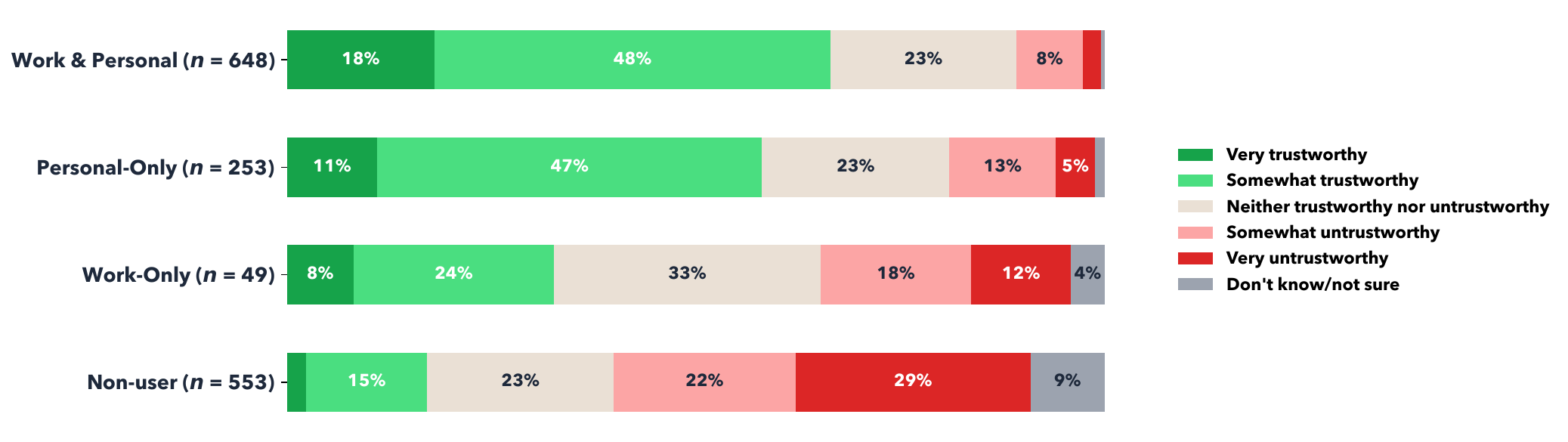}
\caption{Perceived trustworthiness of GenAI responses, split by work and personal use/non-use.}
\label{fig:trust-split}
\end{figure*}

\section{Findings}
\begin{table}[t!]
\centering
\small
\begin{tabular}{llc}
\toprule
Odds for \textbf{Increased}         & \textbf{Odds} \\
{\textbf{Trust in GenAI Outputs}}  & \textbf{Ratio}  & \textbf{\greyT{95\% CI}}  \\  \midrule
\multicolumn{3}{l}{\textbf{\textit{GenAI Use}}} \\
\quad Work-Only & 0.75 & \greyT{[0.33, 1.69]} \\
\quad Personal-Only & \tealT{2.30}*** & \greyT{[1.46, 3.62]} \\
\quad Work \& Personal & 1.49 & \greyT{[0.99, 2.25]} \\  \midrule
\multicolumn{3}{l}{\textbf{\textit{Familiarity with GenAI}}} \\
\quad A little familiar & 2.06 & \greyT{[0.98, 4.31]} \\
\quad Somewhat familiar & \tealT{2.17}* & \greyT{[1.02, 4.63]} \\
\quad Very familiar & \tealT{2.34}* & \greyT{[1.04, 5.24]} \\  \midrule
\multicolumn{3}{l}{\textbf{\textit{GenAI Concern}}} \\
\quad A little concerned & 0.59 & \greyT{[0.34, 1.02]} \\
\quad Somewhat concerned & \pinkT{0.36}*** & \greyT{[0.21, 0.61]} \\
\quad Very/extremely concerned & \pinkT{0.26}*** & \greyT{[0.15, 0.44]} \\  \midrule
\multicolumn{3}{l}{\textbf{\textit{GenAI Excitement}}} \\
\quad A little excited & \tealT{2.67}*** & \greyT{[1.70, 4.18]} \\
\quad Somewhat excited & \tealT{5.05}*** & \greyT{[3.18, 8.02]} \\ 
\quad Very/extremely excited & \tealT{9.38}*** & \greyT{[5.37, 16.41]} \\  \midrule
\multicolumn{3}{l}{\textbf{\textit{GenAI Usefulness}}} \\
\quad A little useful & \tealT{3.66}** & \greyT{[1.45, 9.29]} \\
\quad Somewhat useful & \tealT{9.67}*** & \greyT{[3.90, 23.99]} \\
\quad Very/extremely useful & \tealT{21.17}*** & \greyT{[8.32, 53.84]} \\
\bottomrule
\end{tabular}
\caption{Increased and decreased odds ratios of how likely a respondent is to have high trust in GenAI results. \pinkT{Pink odds ($<$1)} mean a respondent is less likely to have high trust in GenAI outputs than baseline; \tealT{teal odds ($>$1)} mean a respondent is more likely to have high trust in GenAI results than baseline. \emph{Note:} McFadden's Pseudo-$R^2$ = 0.412. *$p$ $<$ 0.05, **$p$ $<$ 0.01, ***$p$ $<$ 0.001. Baseline reference categories: Non-Use, Not Familiar/Never heard of it until now, Not Excited, Not Concerned, Not Useful.}
\label{tab:trust-odds}
\end{table}

\begin{table}[t!]
\centering
\small
\begin{tabular}{llc}
\toprule
Odds for \textbf{Increased}         & \textbf{Odds} \\
{\textbf{Likelihood of GenAI Mistakes}}  & \textbf{Ratio}  & \textbf{\greyT{95\% CI}}  \\  \midrule
\multicolumn{3}{l}{\textbf{\textit{GenAI Use}}} \\
\quad Work-Only & 1.00 & \greyT{[0.49, 2.01]} \\
\quad Personal-Only & 0.71 & \greyT{[0.46, 1.11]} \\
\quad Work \& Personal & 0.92 & \greyT{[0.62, 1.37]} \\  \midrule
\multicolumn{3}{l}{\textbf{\textit{Familiarity with GenAI}}} \\
\quad A little familiar & 0.84 & \greyT{[0.54, 1.30]} \\
\quad Somewhat familiar & \pinkT{1.69}* & \greyT{[1.06, 2.69]} \\
\quad Very familiar & \pinkT{2.64}*** & \greyT{[1.55, 4.48]} \\  \midrule
\multicolumn{3}{l}{\textbf{\textit{GenAI Concern}}} \\
\quad A little concerned & 1.05 & \greyT{[0.63, 1.76]} \\
\quad Somewhat concerned & \pinkT{2.06}** & \greyT{[1.27, 3.35]} \\
\quad Very/extremely concerned & \pinkT{4.55}*** & \greyT{[2.91, 7.11]} \\  \midrule
\multicolumn{3}{l}{\textbf{\textit{GenAI Excitement}}} \\
\quad A little excited & \tealT{0.56}** & \greyT{[0.37, 0.85]} \\
\quad Somewhat excited & \tealT{0.49}** & \greyT{[0.31, 0.79]} \\
\quad Very/extremely excited & 0.80 & \greyT{[0.46, 1.40]} \\  \midrule
\multicolumn{3}{l}{\textbf{\textit{GenAI Usefulness}}} \\
\quad A little useful & 0.75 & \greyT{[0.50, 1.11]} \\
\quad Somewhat useful & \tealT{0.48}** & \greyT{[0.30, 0.76]} \\
\quad Very/extremely useful & \tealT{0.26}*** & \greyT{[0.15, 0.45]} \\ 
\bottomrule
\end{tabular}
\caption{Increased and decreased odds ratios of how likely a respondent is to believe GenAI makes mistakes. \pinkT{Pink odds ($>$1)} mean a respondent is more likely to believe GenAI often/very often makes mistakes; \tealT{teal odds ($<$1)} mean a respondent is more likely to believe that GenAI does \emph{not} often/very often make mistakes. McFadden's Pseudo-$R^2$ = 0.159. *$p$ $<$ 0.05, **$p$ $<$ 0.01, ***$p$ $<$ 0.001. Baseline reference categories: Non-Use, Not Familiar/Never heard of it until now, Not Excited, Not Concerned, Not Useful.}
\label{tab:mistake-odds}
\end{table}

While our primary focus is on GenAI mistakes, we begin with a summary of respondents' general attitudes and use to contextualize our findings in a rapidly changing landscape.

\paragraph{Familiarity with GenAI.} To establish a baseline, we first asked respondents about their familiarity with GenAI (Q1). As reported above, 11\% ($n$ = 163) of respondents were unfamiliar. The vast majority were \emph{familiar} (21\% very familiar, 38\% somewhat familiar, and 31\% a little familiar). For details, see the Appendix, Table 5.

\paragraph{GenAI Use.} We asked familiar respondents how frequently they use GenAI for personal and work tasks (Q7). Across our sample, we again have 11\% of respondents who are unfamiliar, an additional 26\% who do not use GenAI in any context, and the remaining 63\% use GenAI at least rarely in work, personal, or both contexts. More than 5x the number of respondents reported using GenAI in personal tasks but not work tasks, than the reverse. See Table~\ref{tab:use}.

\paragraph{Affect Towards GenAI.} Early in the survey, we asked respondents an open-ended question about their feelings or emotions regarding GenAI (Q2). At a high level, responses bear strong similarity to findings reported for AI prior to the advent of GenAI, e.g., respondents often expressed feeling GenAI is exciting, useful, and/or concerning \cite{kelley2021}. At the same time, certain concerns appear amplified or newly appearing, such as job loss; humans becoming lazy; misinformation (e.g., fake videos/images/news); plagiarism/stealing from artists; and the environment/water use.

\begin{quote}
\emph{``It makes me nervous but also excited about what it can do and how much help it can be. The future is up in the air.''}\footnote{We use verbatim responses throughout without correcting typographic or grammatical errors, but with occasional marked elisions. A few have been translated from Spanish to English.}
\end{quote}

\begin{quote}
\emph{``Excitement for its potential applications in making life easier. Fear that it may lead to misinformation, that it may replace jobs, or result in people becoming lazier and stupider.''}
\end{quote}

At this point in the survey, we had not yet mentioned mistakes. However, responses included some spontaneous mentions of GenAI mistakes, foreshadowing respondent concerns about accuracy which we discuss further below.

\begin{quote}
\emph{``Don't really trust it very much. Often inaccurate.''}
\end{quote}

\begin{quote}
\emph{``Excited but cautious. Appreciative of the abilities of AI but still feel the need to verify the results.''}
\end{quote}

\begin{quote}
\emph{``I worry about Generative AI being so creative that it's not as accurate as it should be...''}
\end{quote}

\begin{quote}
\emph{``I feel that it is helpful and can be a time saver, but am suspicious of the results''}
\end{quote}

\begin{quote}
\emph{``... My job relies on fact, therefore I have little use for AI.''}
\end{quote}

We then asked closed form questions about excitement (Q4), concern (Q5), and usefulness (Q6). For details, see the Appendix, Table 5.

\paragraph{Trustworthiness of GenAI.} Before we asked directly about GenAI mistakes, we asked respondents how trustworthy they feel results generated by AI are (Q9). More respondents report results are trustworthy than untrustworthy, but only 11\% find results very trustworthy. The other 89\% report some form of limited trust, distrust, or aren't sure, with non-users reporting the least trust. See Figures~\ref{fig:trust} and~\ref{fig:trust-split}.

We also review drivers of trust in GenAI results. Excitement about GenAI and belief that GenAI is useful most strongly increase trust in GenAI results (with a respondent who believes GenAI will be very/extremely useful having 21.17x higher odds of trusting GenAI results than someone who believes GenAI will not be useful). Concern with GenAI makes a respondent statistically more likely to distrust GenAI results than one without concern. While familiarity does increase the expectation of trust in GenAI results, these effects are smaller than the others. See Table~\ref{tab:trust-odds}.

\begin{figure*}[t]
\centering
\includegraphics[width=0.9\textwidth]{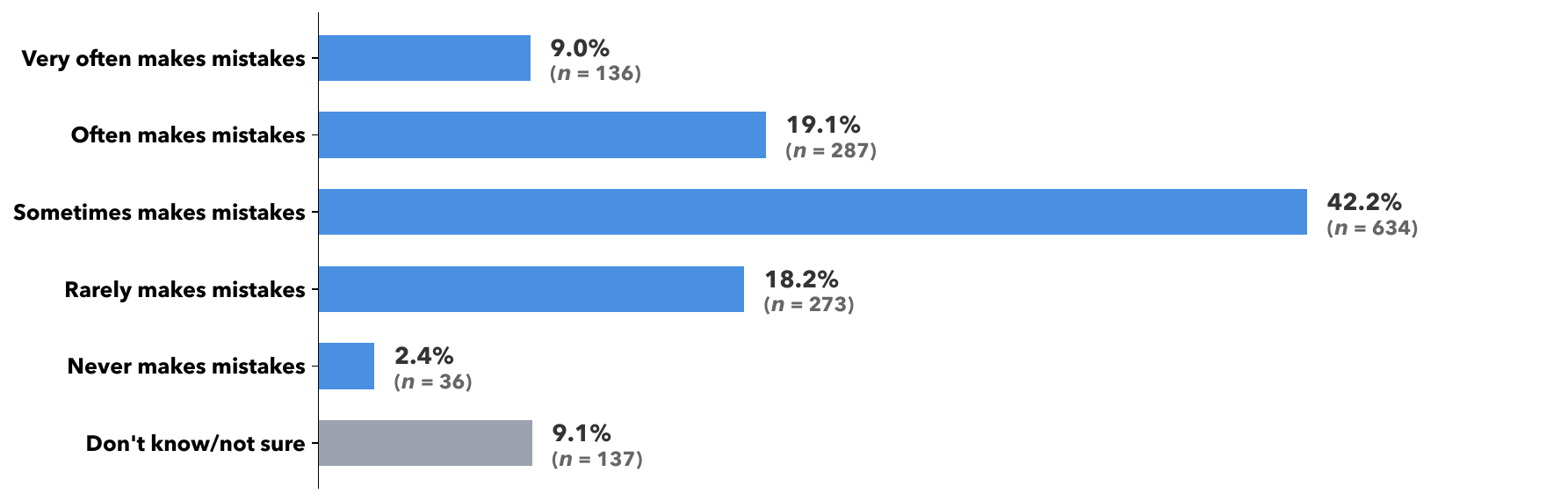}
\caption{Perceived frequency of GenAI making mistakes (Q10).}
\label{fig:mistake-freq}
\end{figure*}

\begin{figure*}[t]
\centering
\includegraphics[width=0.9\textwidth]{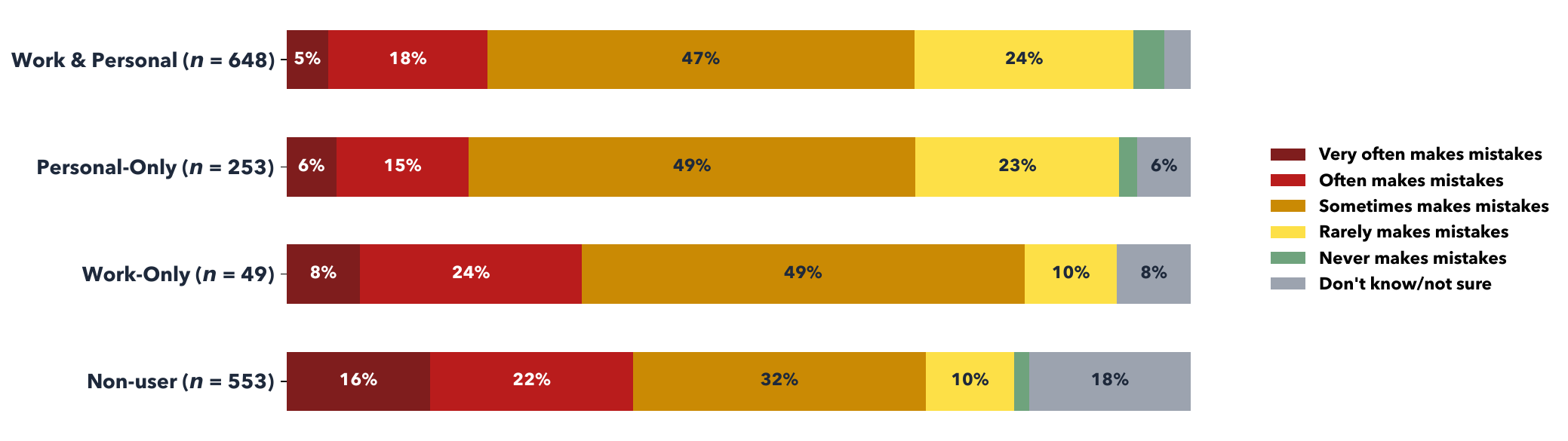}
\caption{Perceived frequency of GenAI making mistakes, split by work and personal use/non-use.}
\label{fig:mistake-freq-split}
\end{figure*}

\subsection{Occurrence of GenAI Mistakes}

For our first research question, we asked all respondents how often they think GenAI tools make mistakes (Q10). Only 2\% ($n$ = 36) said they never make mistakes, with another 18\% ($n$ = 273) saying they rarely make mistakes. In total 70\% ($n$ = 1057), a sizeable majority, said GenAI sometimes, often, or very often makes mistakes. See Figure~\ref{fig:mistake-freq}.

We also see that GenAI users and non-users are broadly aware of mistakes at somewhat similar rates (see Figure~\ref{fig:mistake-freq-split}), an observation supported by reviewing the odds of respondents believing GenAI is often/very often likely to make mistakes, where we see no statistical effect based on use. However, respondents who report they are very familiar with GenAI are 2.64 times more likely to say GenAI often/very often makes mistakes than those who are not familiar; and respondents who report they are very/extremely concerned by GenAI are 4.55 times more likely to say GenAI often/very often makes mistakes than those who are not. We see opposite, though slightly less strong, results for those who are excited about GenAI or believe GenAI is useful, though respondents who say GenAI is very/extremely useful are nearly 4 times \emph{less} likely to state GenAI often/very often makes mistakes than respondents who believe GenAI is not useful. See Table~\ref{tab:mistake-odds}.

\begin{figure*}[t]
\centering
\includegraphics[width=0.95\textwidth]{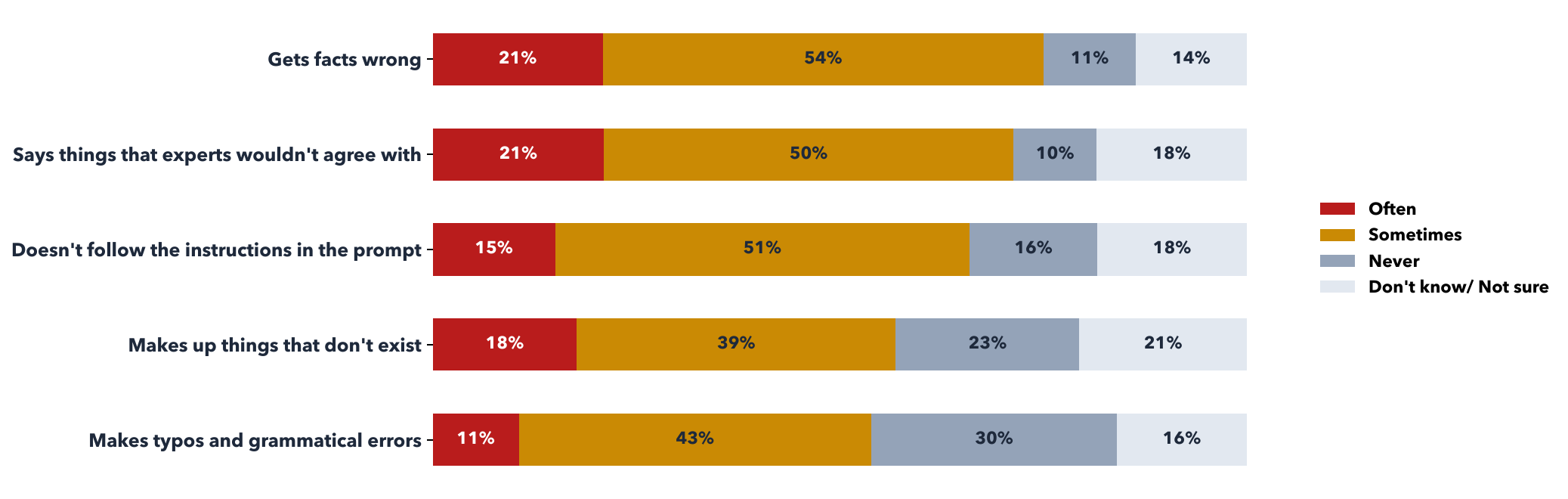}
\caption{Perceived frequency of various GenAI mistakes (Q11).}
\label{fig:mistake-types}
\end{figure*}

\subsection{Types and Frequency of GenAI Mistakes}

We now turn to our second research question, for which we provided a list of five common types of GenAI mistakes and asked respondents how frequently they occur (Q11). Across all five, a majority of respondents said that GenAI often or sometimes makes each type. Respondents thought ``Gets facts wrong'' and ``Says things that experts wouldn't agree with'' were the two most common types of errors, with 21\% of respondents saying these things often happen, and only 11\% and 10\%, respectively, saying they never do. 30\% of respondents said they thought ``Typos and grammatical errors'' never happen (though 11\% thought they happen often, and 43\% thought they happen sometimes). See Figure~\ref{fig:mistake-types}.

\subsection{Checking GenAI Mistakes}

In this section, we report responses to an open-ended question about whether respondents check results from GenAI, and if so, why and how (Q12).

\paragraph{Do people check results from GenAI, and if so, how often?} 75\% ($n$ = 669) of those responding reported checking GenAI results at least sometimes,\footnote{We received 951 responses to this question, and excluded 61 as not relevant, leaving us with a total of 890 valid responses.} with 28\% ($n$ = 250) broadly confirming ``yes'' they check results and an additional 13\% ($n$ = 117) saying they ``always'' check results.

\begin{quote}
\emph{``I always check results. I don't want to blindly rely on AI responses''}
\end{quote}

\begin{quote}
\emph{``Yes I have to check the results. The answer isn't always right. It makes it a little less trustworthy. I check by doing a google search.''}
\end{quote}

\begin{quote}
\emph{``I always check if I'm not sure. If I know enough about the subject that I'm confident in my ability to know if the information is accurate, then I might not double check. Otherwise, I always check by doing searches on one of the search engines or otherwise looking up the information directly. I would never do something based on what generative AI tells me unless I've verified the information first.''}
\end{quote}

\noindent By contrast, only 24\% ($n$ = 216) reported checking GenAI results rarely or not at all, including 18\% ($n$ = 164) saying broadly ``no,'' they do not check results.

\paragraph{Why do people check results from GenAI?} Respondents reported a number of reasons to check GenAI results. Many reported they check ``to make sure'' or ``just in case,'' with a few sharing that everything should be double-checked whether from a human or non-human source. Many said they check because they know GenAI makes mistakes or they do not trust GenAI. Sometimes this mistrust was a general sentiment while in other cases respondents called out specific limitations of GenAI, noted that it is not yet trustworthy because it is a new technology, or said they had learned to mistrust it due to personal experience with inaccurate results.

\begin{quote}
\emph{``I always check the results of work given to me by generative AI. I do not fully trust it and have personally seen it make mistakes before.''}
\end{quote}

\begin{quote}
\emph{``Oh yes, I have to check everything because I find so many mistakes in very simple prompts''}
\end{quote}

\begin{quote}
\emph{``I always check the results. I have been given references that didn't exist. I have been told things that I knew were inaccurate...''}
\end{quote}

\begin{quote}
\emph{``Yes. Because I've come across obvious errors so it's hard to completely trust responses''}
\end{quote}

Some (11\%, $n$ = 99) said they check if results look suspicious, ``absurd,'' or contradict their prior knowledge. Others said they check if the topic is important or work-related, or if sharing incorrect results might embarrass them.

\begin{quote}
\emph{``Always as I am responsible for the results.''}
\end{quote}

\begin{quote}
\emph{``Yes because I don't want to have the wrong thing down which then makes me look stupid when other people read my work''}
\end{quote}

\begin{quote}
\emph{``Yes I am required to check all the responses per my company AI use policies. I will use another tool to check the responses that seem suspect.''}
\end{quote}

\begin{quote}
\emph{``Yes, I check the results because I do not want to send something wrong out to a client.''}
\end{quote}

\paragraph{Why do people \emph{not} check results from GenAI?} Respondents also reported reasons \emph{not} to check GenAI results. Some said they do not check for unimportant tasks. Some went further, saying that they do not need to check in general because they do not use GenAI for anything important.

\begin{quote}
\emph{``Whenever I've used it, it's been primarily for entertainment purposes, so I don't take it too seriously---nor do I feel the need to verify what it generates for me.''}
\end{quote}

\begin{quote}
\emph{``I rarely check. None of my activities using AI are of sufficient importance to spend time checking the results. If I suspect something is amiss, I will do a normal search on Google or Bing or generally on the internet.''}
\end{quote}

Others said it is unnecessary to check because they trust GenAI.\footnote{In some cases, a reason \emph{not to} check was naturally the inverse of a reason \emph{to} check. For example, trust was a reason not to check while lack of trust was a reason to check.} Some reported their trust stemmed from personal experience, or their understanding of GenAI's data sources. Others shared that they initially checked GenAI results, but have come to trust GenAI over time and now feel less need to check.

\begin{quote}
\emph{``I hardly ever check the results from Ai because it's usually pretty accurate.''}
\end{quote}

\begin{quote}
\emph{``I never check results. I beleive AI tools because they provide responses from existing data available on the web.''}
\end{quote}

\begin{quote}
\emph{``No I trust it because it has access to the entire internet''}
\end{quote}

\begin{quote}
\emph{``At first I would check the results of AI. Now that I have used it for an extended period of time I have become more trusting of its responses.''}
\end{quote}

Some said that they do not check because it would be a poor use of time and/or checking would defeat the purpose of using GenAI, particularly since they are often using GenAI to be more efficient.

\begin{quote}
\emph{``I don't check the results because I'm using AI to be quick and that's an extra step.''}
\end{quote}

\begin{quote}
\emph{``I actually have never checked the results out of laziness.''}
\end{quote}

\paragraph{How do people check GenAI results?} Respondents are motivated to check GenAI results for accuracy, for example, ``to be really certain that I have my facts right'' or because they want to know ``the truth.'' A few offered other reasons such as checking that the information is current, or a broad goal of checking to make sure the results are ``okay'' and of sufficiently high quality. In order to achieve these goals, respondents would often check results by cross-referencing with other sources (25\%; $n$ = 226). Search engines and websites were a common resource, including some respondents reporting they check the sources listed in GenAI results. A few reported comparing results from multiple GenAI tools. Some respondents also emphasized the value of comparing with reputable sources or conferring with human experts.

\begin{quote}
\emph{``Yes. AI is notoriously inaccurate, getting information from Reddit and other online forms where anyone can say anything. I check with a Google search and look for reputable sources that echo what AI is saying.''}
\end{quote}

\begin{quote}
\emph{``I like to check the results to make sure that the information is correct. I check other websites such as google, etc. and see if the information is consistent.''}
\end{quote}

\begin{quote}
\emph{``I often check the links it provides to make sure it's right when I'm using the information for a consequential task.''}
\end{quote}

\begin{quote}
\emph{``I normally read through the results and if something seems off I ask someone about it or search it up.''}
\end{quote}

Beyond checking for accuracy, others reported checking results for relevance, as a number of respondents reported experiences with misinterpreted prompts and/or irrelevant results. Others said they manually review presentation aspects such as grammar, tone, and appropriateness.

\subsection{Explanations and In-Product Statements}

During the highlighter exercise, respondents read and assessed hypothetical explanations (longer text, such as might appear in a blog post) (Q14); responded to an open-ended question about whether they would want any additional information beyond that included in the explanations (Q14b); and finally read and assessed hypothetical in-product statements (shorter text, such as might appear next to GenAI output) (Q15).

\begin{figure*}[p!]
\centering
\includegraphics[width=0.98\textwidth]{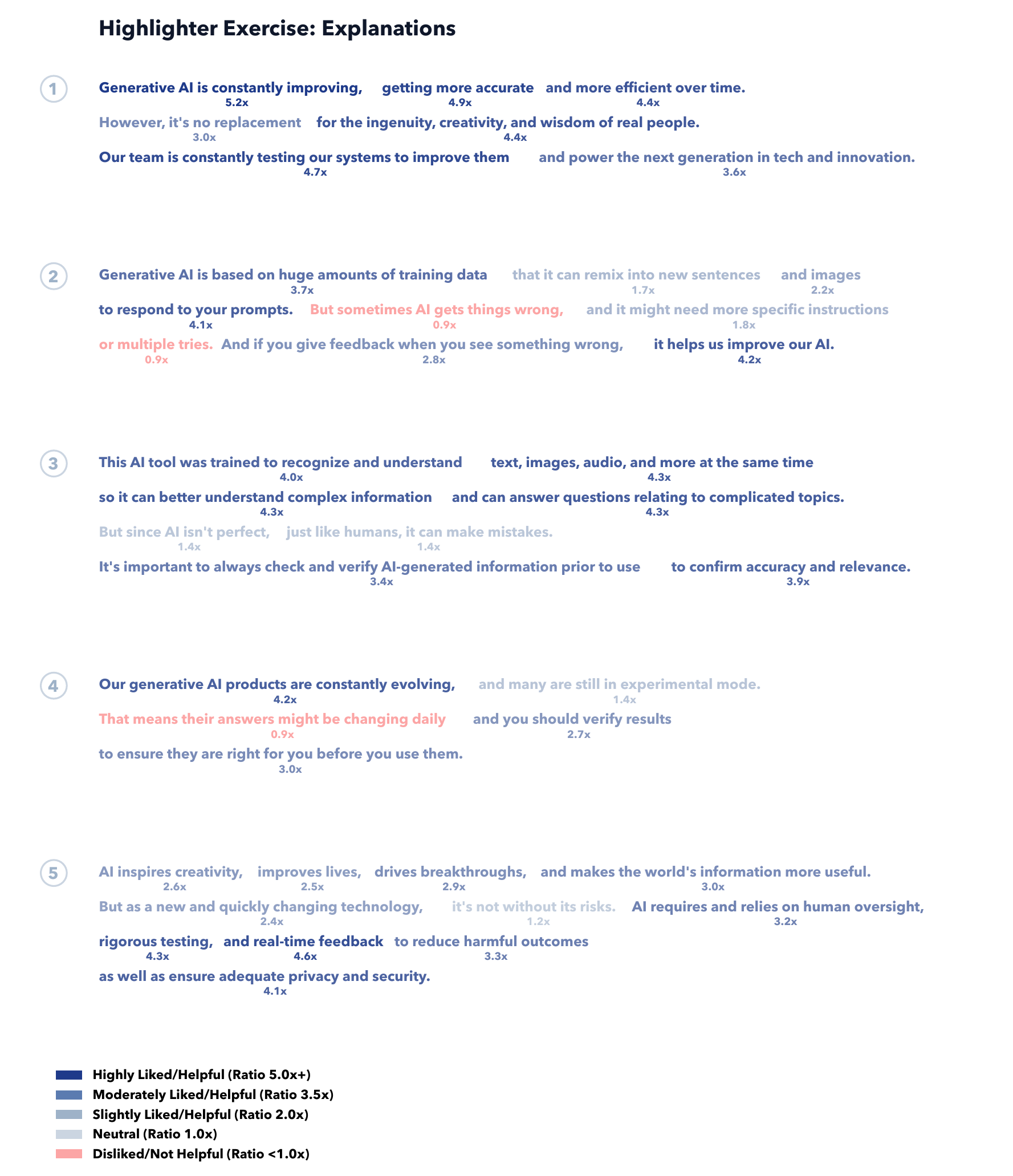}
\caption{Highlighter exercise results on hypothetical text explanations, similar to what might appear in a blog post or an About page. Color indicates the ratio of those who liked/found helpful a phrase versus disliked/found it not helpful (Q14).}
\label{fig:highlighter-explanations}
\end{figure*}

\begin{figure*}[t!]
\centering
\includegraphics[width=0.98\textwidth]{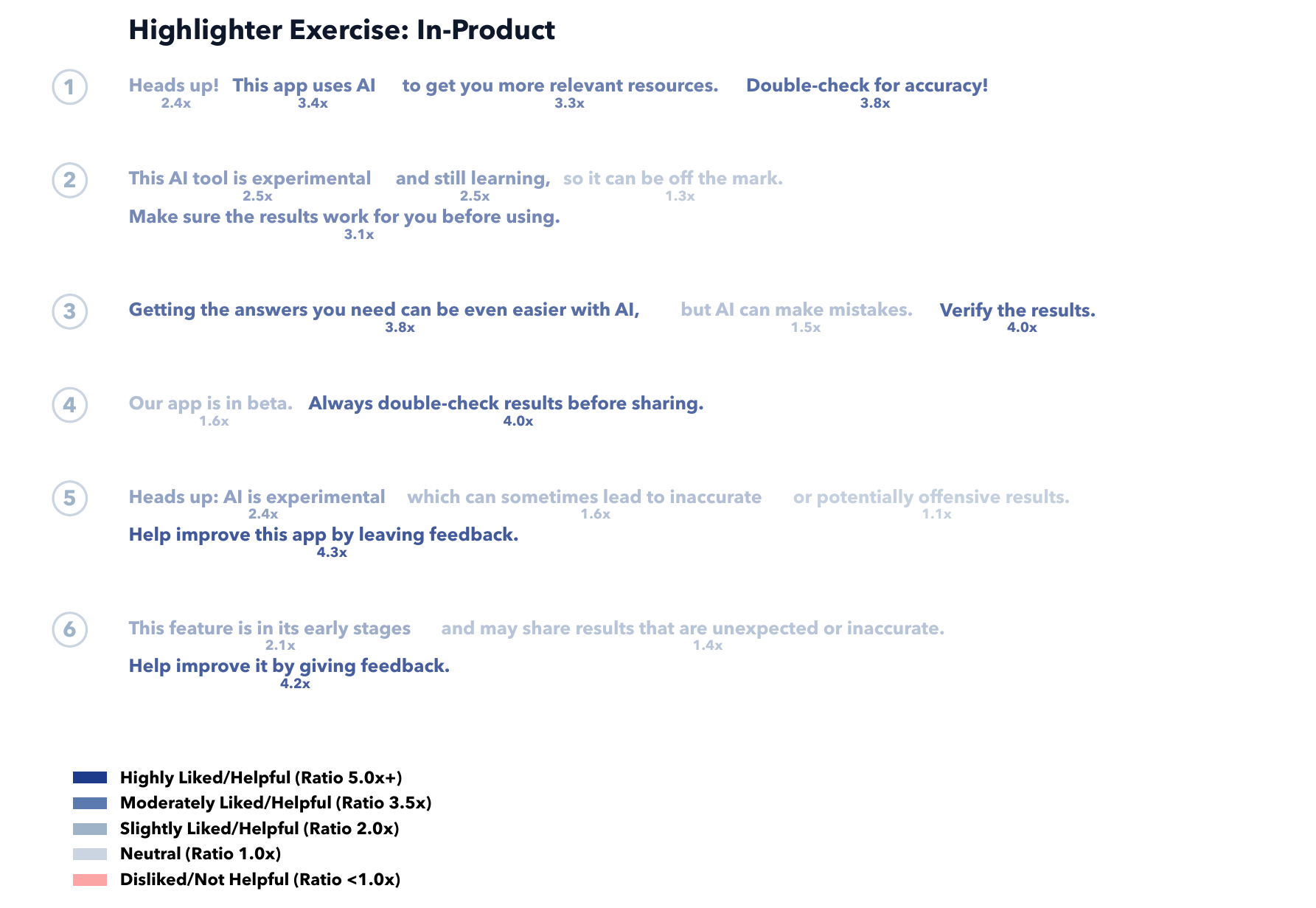}
\caption{Highlighter exercise results on hypothetical in-product disclosures. Color indicates the ratio of those who liked/found helpful a phrase versus disliked/found it not helpful (Q15).}
\label{fig:highlighter-inproduct}
\end{figure*}

For the explanations, respondents gravitated toward text that emphasized constantly improving GenAI and user feedback, while also highlighting that results need to be checked. Respondents were less favorable toward descriptions that focused on GenAI getting things wrong or making mistakes, and particularly unfavorable toward text about GenAI giving inconsistent results. For the in-product statements, respondents gravitated toward reminders to check results but did not like or find helpful reminders that GenAI makes mistakes or can give inaccurate information. See Figures~\ref{fig:highlighter-explanations} and~\ref{fig:highlighter-inproduct}.

In the open-ended question, many respondents shared they do not need additional information beyond that provided in the explanations, and some expressed frustration with current limitations of GenAI. At the same time, many others did want to learn more, particularly about the following: \emph{verification procedures} (what specifically should users do to check results?); \emph{accuracy} (specifically how accurate is GenAI, and will it continue to make mistakes ``indefinitely?''); \emph{inputs and training} (where does GenAI get its information?); and \emph{societal impact} (e.g., the environment).

\begin{quote}
\emph{``I would like to know how they want us to double check. Also, explain environmental consequences of using ai.''}
\end{quote}

\begin{quote}
\emph{``having to check the results seems to defeat the purpose of AI. Suggestions on how to do that quickly and easily would be helpful.''}
\end{quote}

\begin{quote}
\emph{``I would want to know how frequently AI is wrong.''}
\end{quote}

\begin{quote}
\emph{``What categories are more reliable? What \% of errors are found?''}
\end{quote}

\begin{quote}
\emph{``I will like it more when it can be almost 100\% accurate in most cases''}
\end{quote}

\section{Discussion and Conclusions}

We discuss implications for the design of GenAI disclosures, and then turn to opportunities for future work.

\subsection{Guidance for Disclosures of GenAI Mistakes}

Based on our results, we highlight five insights to help users grasp the limitations of GenAI and consider possible actions to take. We hope such guidance may be useful to developers of GenAI products and services, as well as regulators and civil society groups considering policy implications and best practices for disclosures and disclaimers of GenAI mistakes.

\paragraph{Avoid vague, repeated warnings about GenAI mistakes.} Respondents, largely already aware of GenAI's potential to err, disfavored general language about mistakes and inaccuracy. Phrases like ``AI can make mistakes'' or ``inaccurate'' consistently received lower marks. While acknowledging imperfection is important, negative reactions to phrases like ``AI gets things wrong'' and ``it's not without its risks'' suggest that cautionary notes should be balanced and framed constructively. Although it remains important to clearly state that GenAI can make mistakes, it appears that given current public understanding, at least among regular or heavy users, it is not necessary to repeatedly remind the user of this. If the same in-product statement about mistakes is seen by users every time they prompt a GenAI tool, this repeated messaging may be unhelpful or disliked. (Note that new users or those just beginning to explore GenAI may need such reminders more often.)

\paragraph{Provide clear, actionable advice for checking results.} Respondents responded positively to language that directly encourages them to ``double-check for accuracy,'' ``verify results,'' or ``make sure the results work for you before using,'' suggesting that users value specific, empowering guidance rather than vague warnings. At the same time, some respondents wanted more detailed guidance on how they should perform these checks.

\paragraph{Remind users that they can give feedback, and that feedback helps improve AI.} Respondents responded positively to statements that for many GenAI tools they can provide feedback, flagging results that they believe may be wrong, harmful, or questionable. They also responded positively to text about feedback helping improve GenAI tools.

\paragraph{Describe the processes that make the product/service safe and accurate.} Respondents reacted positively to information about how tools are made more accurate and responsible, for example, ``huge amounts of training data,'' ``rigorous testing,'' and ``ensure adequate privacy and security.'' Respondents here again wanted more detailed information about these assurances, particularly regarding which data is used for training and whether the creators of that data consented to its inclusion in training. They also expressed interest in other measures, such as the steps companies are taking to lower the environmental costs of the models.

\paragraph{Describe improvement to the models over time.} As GenAI tools are rapidly evolving, it is useful to remind users that GenAI continues to improve. Phrases like ``constantly improving,'' ``getting more accurate,'' and ``evolving'' were consistently marked as liked or helpful. Currently, it is useful to explain to the general public that a lot of effort is going into improving GenAI products and the underlying models.

\subsection{Future Work}

Our results demonstrate high public awareness of GenAI mistakes among both users and non-users of GenAI. These results suggest substantial opportunities to further investigate the public's attitudes and behaviors. Moreover, as technologies such as agentic AI evolve, it will be valuable to continue to evaluate public opinion. Further, as models become more advanced and mistakes become fewer or more subtle, user inclination to check results may change.

While we are cautious of putting too much burden on users to ensure accuracy, our findings suggest opportunities to develop effective messaging to educate users about GenAI mistakes, and to support human-AI verification. Respondents' interest in providing feedback offers opportunities for transparency about the use of such feedback, consistent with best practices outlined in explainability rubrics \cite{kelley2023}. Respondents' interest in accuracy and their commitment to checking results is encouraging for the development of verification support tools \cite{gordon2023,nahar2024,zhou2026,laban2024}. High quality verification tools are especially important when considering the limits of users' ability to check results. For example, some users reported looking for ``suspicious'' results as the trigger to verify; this may be problematic if it causes users to skip verification when the output contains only difficult-to-detect errors they do not spot.

\section{Acknowledgments}

We thank Nina Trach, Emily Kostic, Laurie Keith, Hannah Lushin, and Tyler Hansen of The Ad Council for their valuable contributions to this work. We thank C+R for their excellent work fielding the survey. We are grateful to Alexis Reiter, Amanda Storey, and Ashley Walker of Google for supporting this work.

\bibliography{aaai2026}

@article{annapureddy2025,
  author  = {Annapureddy, Raghav and Fornaroli, Alessandro and Gatica-Perez, Daniel},
  title   = {Generative {AI} Literacy: Twelve Defining Competencies},
  journal = {Digital Government: Research and Practice},
  volume  = {6},
  number  = {1},
  year    = {2025}
}

@book{beyer1997,
  author    = {Beyer, Hugh and Holtzblatt, Karen},
  title     = {Contextual Design: Defining Customer-Centered Systems},
  year      = {1997},
  publisher = {Elsevier}
}

@article{bick2026,
  author  = {Bick, Alexander and Blandin, Adam and Deming, David J.},
  title   = {The Rapid Adoption of Generative {AI}},
  journal = {Management Science},
  year    = {2026},
}

@article{bohannon2023,
  author  = {Bohannon, Molly},
  title   = {Lawyer Used {ChatGPT} In Court---And Cited Fake Cases. A Judge Is Considering Sanctions},
  journal = {Forbes},
  year    = {2023},
}

@techreport{carmichael2025,
  author      = {Carmichael, Mallory},
  title       = {The Ipsos {AI} Monitor 2025},
  institution = {Ipsos},
  type        = {Report},
  year        = {2025},
}

@techreport{chatterji2025,
  author      = {Chatterji, Aaron and Cunningham, Tom and Deming, David J. and Hitzig, Zoe and Ong, Christopher and Shan, Carl Ye and Wadman, Kevin},
  title       = {How People Use {ChatGPT}},
  institution = {National Bureau of Economic Research},
  type        = {Working Paper},
  number      = {34255},
  year        = {2025},
}

@misc{eddy2025,
  author       = {Eddy, Katherine},
  title        = {Americans Have Mixed Feelings about {AI} Summaries in Search Results},
  howpublished = {Short Read, Pew Research Center},
  year         = {2025},
}

@misc{feldmann2025,
  author       = { Derrick Feldmann and Colleen Thompson-Kuhn and Nina Trach and Emily Kostic and Laurie Keith and Hannah Lushin and Tyler Hansen and Reena Jana and Patrick Gage Kelley and Allison Woodruff},
  title        = {Calibrating Trustworthiness in {GenAI}: Perceptions of generative {AI} results and messaging that can help the American public assess trustworthiness in {GenAI}},
  howpublished = {Ad Council Research Institute},
  year         = {2025}
}

@article{goldberg2026,
  author  = {Goldberg, Michelle},
  title   = {The Generation That Grew Up With {A.I.} Hates It},
  journal = {The New York Times},
  year    = {2026},
}

@misc{gordon2023,
  author        = {Gordon, Andrew D. and Negreanu, Cristian and Cambronero, Jos{\'e} and Chakravarthy, Raj and Drosos, Ian and Fang, Hao and Mitra, Bhaskar and Richardson, Hannah and Sarkar, Advait and Simmons, Steve and Williams, Jack and Zorn, Benjamin},
  title         = {Co-audit: tools to help humans double-check {AI}-generated content},
  year          = {2023},
  eprint        = {2310.01297},
  archivePrefix = {arXiv},
}

@inproceedings{gu2024,
  author    = {Gu, Ken and Shang, Ruoxi and Althoff, Tim and Wang, Chenglong and Drucker, Steven M.},
  title     = {How Do Analysts Understand and Verify {AI}-Assisted Data Analyses?},
  booktitle = {Proceedings of the 2024 CHI Conference on Human Factors in Computing Systems},
  series    = {CHI '24},
  year      = {2024},
  address   = {New York, NY, USA},
  publisher = {Association for Computing Machinery},
}

@article{hinkin1998,
  author  = {Hinkin, Timothy R.},
  title   = {A brief tutorial on the development of measures for use in survey questionnaires},
  journal = {Organizational Research Methods},
  volume  = {1},
  number  = {1},
  pages   = {104--121},
  year    = {1998},
}

@article{holbrook2003,
  author  = {Holbrook, Allyson L. and Green, Melanie C. and Krosnick, Jon A.},
  title   = {Telephone versus face-to-face interviewing of national probability samples with long questionnaires: Comparisons of respondent satisficing and social desirability response bias},
  journal = {Public Opinion Quarterly},
  volume  = {67},
  number  = {1},
  pages   = {79--125},
  year    = {2003},
}

@misc{ipsos2026a,
  author       = {{Ipsos}},
  title        = {Americans think they need to keep up with {AI}, but {AI} needs to slow down},
  howpublished = {Ipsos Consumer Tracker},
  year         = {2026},
}

@misc{ipsos2026b,
  author       = {{Ipsos}},
  title        = {Generative {AI} use is getting more mainstream in America},
  howpublished = {Ipsos Consumer Tracker},
  year         = {2026},
}

@article{ji2023,
  author  = {Ji, Ziwei and Lee, Nayeon and Frieske, Rita and Yu, Tiezheng and Su, Dan and Xu, Yan and Ishii, Etsuko and Bang, Ye Jin and Madotto, Andrea and Fung, Pascale},
  title   = {Survey of Hallucination in Natural Language Generation},
  journal = {ACM Computing Surveys},
  volume  = {55},
  number  = {12},
  year    = {2023},
}

@article{kelley2023,
  author  = {Kelley, Patrick Gage and Woodruff, Allison},
  title   = {Advancing Explainability Through {AI} Literacy and Design Resources},
  journal = {Interactions},
  volume  = {30},
  number  = {5},
  pages   = {34--38},
  year    = {2023},
}

@inproceedings{kelley2021,
  author    = {Kelley, Patrick Gage and Yang, Yongwei and Heldreth, Courtney and Moessner, Christopher and Sedley, Aaron and Kramm, Andreas and Newman, David T. and Woodruff, Allison},
  title     = {Exciting, Useful, Worrying, Futuristic: Public Perception of Artificial Intelligence in 8 Countries},
  booktitle = {Proceedings of the 2021 AAAI/ACM Conference on AI, Ethics, and Society},
  series    = {AIES '21},
  pages     = {627--637},
  year      = {2021},
  address   = {New York, NY, USA},
  publisher = {Association for Computing Machinery},
}

@techreport{kennedy2025,
  author      = {Kennedy, Brian and Yam, Elisha and Kikuchi, Elizabeth and Pula, Ilana and Fuentes, Julian},
  title       = {How Americans View {AI} and Its Impact on People and Society},
  institution = {Pew Research Center},
  type        = {Report},
  year        = {2025},
}

@inproceedings{laban2024,
  author    = {Laban, Philippe and Vig, Jesse and Hearst, Marti and Xiong, Caiming and Wu, Chien-Sheng},
  title     = {Beyond the Chat: Executable and Verifiable Text-Editing with {LLMs}},
  booktitle = {Proceedings of the 37th Annual ACM Symposium on User Interface Software and Technology},
  series    = {UIST '24},
  year      = {2024},
  address   = {New York, NY, USA},
  publisher = {Association for Computing Machinery},
}

@misc{leanin2026,
  author       = {{Lean In}},
  title        = {Women use {AI} less often at work and get less credit},
  howpublished = {LeanIn.Org},
  year         = {2026},
}

@inproceedings{lee2025,
  author    = {Lee, Hao-Ping (Hank) and Sarkar, Advait and Tankelevitch, Lev and Drosos, Ian and Rintel, Sean and Banks, Richard and Wilson, Nicholas},
  title     = {The Impact of Generative {AI} on Critical Thinking: Self-Reported Reductions in Cognitive Effort and Confidence Effects From a Survey of Knowledge Workers},
  booktitle = {Proceedings of the 2025 CHI Conference on Human Factors in Computing Systems},
  series    = {CHI '25},
  year      = {2025},
  address   = {New York, NY, USA},
  publisher = {Association for Computing Machinery},
}

@misc{li2026,
  author        = {Li, Mingyi and Bickersteth, William and Tang, Nina and Kapoor, Priya and Win, Khine and Zhong, Peiyi and Hong, Jason I. and Cranor, Lorrie Faith and Heidari, Hoda and Shen, Hong},
  title         = {What People See (and Miss) About Generative {AI} Risks: Perceptions of Failures, Risks, and Who Should Address Them},
  year          = {2026},
  eprint        = {2604.22654},
  archivePrefix = {arXiv},
}

@techreport{lin2025,
  author      = {Lin, Luona and Parker, Kim},
  title       = {{U.S.} Workers Are More Worried Than Hopeful About Future {AI} Use in the Workplace},
  institution = {Pew Research Center},
  type        = {Report},
  year        = {2025},
}

@inproceedings{long2020,
  author    = {Long, Duri and Magerko, Brian},
  title     = {What is {AI} Literacy? Competencies and Design Considerations},
  booktitle = {Proceedings of the 2020 CHI Conference on Human Factors in Computing Systems},
  series    = {CHI '20},
  pages     = {1--16},
  year      = {2020},
  address   = {New York, NY, USA},
  publisher = {Association for Computing Machinery},
}

@inproceedings{mahmood2022,
  author    = {Mahmood, Amama and Fung, Jeanie W. and Won, Isabel and Huang, Chien-Ming},
  title     = {Owning Mistakes Sincerely: Strategies for Mitigating {AI} Errors},
  booktitle = {Proceedings of the 2022 CHI Conference on Human Factors in Computing Systems},
  series    = {CHI '22},
  year      = {2022},
  address   = {New York, NY, USA},
  publisher = {Association for Computing Machinery},
}

@inproceedings{mcdonald2019,
  author    = {McDonald, Nora and Schoenebeck, Sarita and Forte, Andrea},
  title     = {Reliability and Inter-rater Reliability in Qualitative Research: Norms and Guidelines for {CSCW} and {HCI} Practice},
  booktitle = {Proceedings of the 22nd ACM Conference on Computer Supported Cooperative Work and Social Computing (CSCW 2019)},
  year      = {2019},
}

@article{mcgregor2021,
  author  = {McGregor, Sean},
  title   = {Preventing Repeated Real World {AI} Failures by Cataloging Incidents: The {AI} Incident Database},
  journal = {Proceedings of the AAAI Conference on Artificial Intelligence},
  volume  = {35},
  number  = {17},
  pages   = {15458--15463},
  year    = {2021},
}

@article{mickle2026,
  author  = {Mickle, Tripp},
  title   = {From Indiana to Idaho, a Backlash Against {A.I.} Gathers Momentum},
  journal = {The New York Times},
  year    = {2026},
}

@misc{mitfuturetech2025,
  author       = {{MIT FutureTech}},
  title        = {The {AI} Risk Repository},
  howpublished = {\url{https://airisk.mit.edu/}},
  note         = {Accessed: 2026-05-19},
  year         = {2025},
}

@article{mueller2025,
  author  = {Mueller, Andreas and Kuester, Sabine and von Janda, Sergej},
  title   = {Socially (un)acceptable errors of {AI}: Consumer perceptions of different {AI}-induced errors},
  journal = {Journal of Business Research},
  volume  = {201},
  pages   = {115673},
  year    = {2025},
}

@inproceedings{nahar2024,
  author    = {Nahar, Mahjabin and Seo, Haeseung and Lee, Eun-Ju and Xiong, Aiping and Lee, Dongwon},
  title     = {Fakes of Varying Shades: How Warning Affects Human Perception and Engagement Regarding {LLM} Hallucinations},
  booktitle = {Proceedings of the Conference on Language Modeling},
  series    = {COLM 2024},
  year      = {2024},
}

@techreport{passi2022,
  author      = {Passi, Samir and Vorvoreanu, Mihaela},
  title       = {Overreliance on {AI}: Literature Review},
  institution = {Microsoft},
  type        = {Technical Report},
  number      = {MSR-TR-2022-12},
  year        = {2022},
}

@inproceedings{perera2026,
  author    = {Perera, Manaswi and Ananthanarayan, Swamy and Goncu, Cagatay and Marriott, Kim},
  title     = {I'm Always a Little Skeptical of It: Verification Practices of Blind Users When Working with Generative {AI} in Spreadsheets},
  booktitle = {Proceedings of the 2026 CHI Conference on Human Factors in Computing Systems},
  series    = {CHI '26},
  year      = {2026},
  address   = {New York, NY, USA},
  publisher = {Association for Computing Machinery},
}

@techreport{poushter2025,
  author      = {Poushter, Jacob and Fagan, Moira and Corichi, Manolo},
  title       = {How People Around the World View {AI}},
  institution = {Pew Research Center},
  type        = {Report},
  year        = {2025},
}

@misc{quinnipiac2026,
  author       = {{Quinnipiac University Poll}},
  title        = {The Age Of Artificial Intelligence: Americans' {AI} Use Increases While Views On It Sour, Quinnipiac University Poll On {AI} Finds; 7 In 10 Think {AI} Will Cut Jobs With Gen {Z} The Most Pessimistic},
  howpublished = {Press Release},
  year         = {2026},
}

@inproceedings{raji2022,
  author    = {Raji, Inioluwa Deborah and Kumar, I. Elizabeth and Horowitz, Aaron and Selbst, Andrew},
  title     = {The Fallacy of {AI} Functionality},
  booktitle = {Proceedings of the 2022 ACM Conference on Fairness, Accountability, and Transparency},
  series    = {FAccT '22},
  pages     = {959--972},
  year      = {2022},
  address   = {New York, NY, USA},
  publisher = {Association for Computing Machinery},
}

@techreport{sajadieh2026,
  author      = {Sajadieh, Sarah and Fattorini, Loredana and Perrault, Raymond and Gil, Yolanda and Parli, Vanessa and Santarlasci, Lauren and Pava, Juan and Maslej, Nestor and Altman, Russell and Brynjolfsson, Erik and Brodley, Cathy and Clark, Jack and Dignum, Virginia and Kumar, Vidushi and Landay, James and Lyons, Terah and Manyika, James and Niebles, Juan Carlos and Shoham, Yoav and Tabassi, Elham and Wald, Roy and Walsh, Toby and Weld, Daniel},
  title       = {The {AI} Index 2026 Annual Report},
  institution = {AI Index Steering Committee, Institute for Human-Centered AI, Stanford University},
  address     = {Stanford, CA},
  type        = {Report},
  year        = {2026},
}

@inproceedings{shelby2023,
  author    = {Shelby, Renee and Rismani, Shalaleh and Henne, Kathryn and Moon, AJung and Rostamzadeh, Negar and Nicholas, Paul and Yilla-Akbari, N'Mah and Gallegos, Jess and Smart, Andrew and Garcia, Emilio and Virk, Gurleen},
  title     = {Sociotechnical Harms of Algorithmic Systems: Scoping a Taxonomy for Harm Reduction},
  booktitle = {Proceedings of the 2023 AAAI/ACM Conference on AI, Ethics, and Society},
  series    = {AIES '23},
  pages     = {723--741},
  year      = {2023},
  address   = {New York, NY, USA},
  publisher = {Association for Computing Machinery},
}

@misc{sidoti2025,
  author       = {Sidoti, Olivia and McClain, Colleen},
  title        = {34\% of {U.S.} adults have used {ChatGPT}, about double the share in 2023},
  howpublished = {Short Read, Pew Research Center},
  year         = {2025},
}

@inproceedings{tankelevitch2024,
  author    = {Tankelevitch, Lev and Kewenig, Viktor and Simkute, Auste and Scott, Ava Elizabeth and Sarkar, Advait and Sellen, Abigail and Rintel, Sean},
  title     = {The Metacognitive Demands and Opportunities of Generative {AI}},
  booktitle = {Proceedings of the 2024 CHI Conference on Human Factors in Computing Systems},
  series    = {CHI '24},
  year      = {2024},
  address   = {New York, NY, USA},
  publisher = {Association for Computing Machinery},
}

@inproceedings{tolsdorf2025,
  author    = {Tolsdorf, Jan and Luo, April Fang and Kodwani, Meghana and Eum, Junghyun and Sharif, Mahmood and Mazurek, Michelle L. and Aviv, Adam J.},
  title     = {Safety perceptions of generative {AI} conversational agents: uncovering perceptual differences in trust, risk, and fairness},
  booktitle = {Proceedings of the Twenty-First USENIX Conference on Usable Privacy and Security},
  series    = {SOUPS '25},
  year      = {2025},
  address   = {USA},
  publisher = {USENIX Association},
}

@inproceedings{zhou2026,
  author    = {Zhou, Ruiyang and Nguyen, Giang and Kharya, Nihal and Nguyen, An and Agarwal, Chirag},
  title     = {Improving Human Verification of {LLM} Reasoning through Interactive Explanation Interfaces},
  booktitle = {Proceedings of the 31st International Conference on Intelligent User Interfaces},
  series    = {IUI '26},
  pages     = {456--473},
  year      = {2026},
  address   = {New York, NY, USA},
  publisher = {Association for Computing Machinery},
}

\end{document}